\documentclass[aps,prl,twocolumn,showpacs,superscriptaddress,a4paper,floatfix]{revtex4}
\usepackage[latin1]{inputenc}
\usepackage{graphicx}
\usepackage{amsmath,amssymb}
\usepackage{hyperref}
\usepackage{datetime}
\usepackage{tikz}
\usepackage{upgreek}

\begin{document}

\title{Connecting strongly correlated superfluids by a quantum point contact}
\author{Dominik Husmann}
\affiliation{Department of Physics, ETH Zurich, 8093 Zurich, Switzerland}
\author{Shun Uchino}
\affiliation{Department of Quantum Matter, Universit\'e de Gen\`eve, CH-1211 Gen\`eve, Switzerland}
\author{Sebastian Krinner}
\affiliation{Department of Physics, ETH Zurich, 8093 Zurich, Switzerland}
\author{Martin Lebrat}
\affiliation{Department of Physics, ETH Zurich, 8093 Zurich, Switzerland}
\author{Thierry Giamarchi}
\affiliation{Department of Quantum Matter, Universit\'e de Gen\`eve, CH-1211 Gen\`eve, Switzerland}
\author{Tilman Esslinger}
\affiliation{Department of Physics, ETH Zurich, 8093 Zurich, Switzerland}
\author{Jean-Philippe Brantut}
\affiliation{Department of Physics, ETH Zurich, 8093 Zurich, Switzerland}
\date{\pdfdate}

\begin{abstract}
Point contacts provide simple connections between macroscopic particle 
reservoirs. In electric circuits, strong links between metals, semiconductors 
or superconductors have applications for fundamental condensed-matter physics 
as well as quantum information processing. However for complex, strongly 
correlated materials, links have been largely restricted to weak tunnel 
junctions. Here we study resonantly interacting Fermi gases connected by a 
tunable, ballistic quantum point contact, finding a non-linear current-bias 
relation. At low temperature, our observations agree quantitatively with a 
theoretical model in which the current originates from multiple Andreev 
reflections. In a wide contact geometry, the competition between 
superfluidity and thermally activated transport leads to a conductance 
minimum. Our system offers a controllable platform for the study of 
mesoscopic devices based on strongly interacting matter.
\end{abstract}

\maketitle

The effect of strong interactions between the constituents of a quantum
many-body system is at the origin of several challenging questions in
physics. Whilst the ground states of strongly interacting systems are
increasingly better understood \cite{Leggett:2006aa}, the properties out of 
equilibrium and at finite temperature often remain puzzling, as these
are determined by the  excitations above the ground state. In laboratory
experiments, strongly interacting systems are found in certain
materials, as well as in quantum fluids and gases
\cite{Leggett:2006aa}. In solid-state systems, a conceptually simple and
clean approach to probe non-equilibrium physics is provided by transport
measurements through the well-defined geometry of a quantum point
contact (QPC) \cite{Post:1994aa,Scheer:1997aa,Fischer:2007aa}. Yet, the 
technical hurdles to realise a controlled QPC between strongly correlated materials 
pose a big challenge. Ultra-cold atomic Fermi gases in the vicinity of a Feshbach
resonance, the so-called unitary regime, provide an alternative route to 
study correlated systems \cite{Zwerger:2011aa}. Superfluidity has been 
established at low temperature \cite{Zwierlein:2005ab}, but the finite-
temperature properties are only partially understood \cite{Nascimbene:2011aa,Ku:2012aa,Sagi:2015aa}, a situation similar to the field of strongly 
correlated materials.

\begin{figure*}[htb]
	\includegraphics{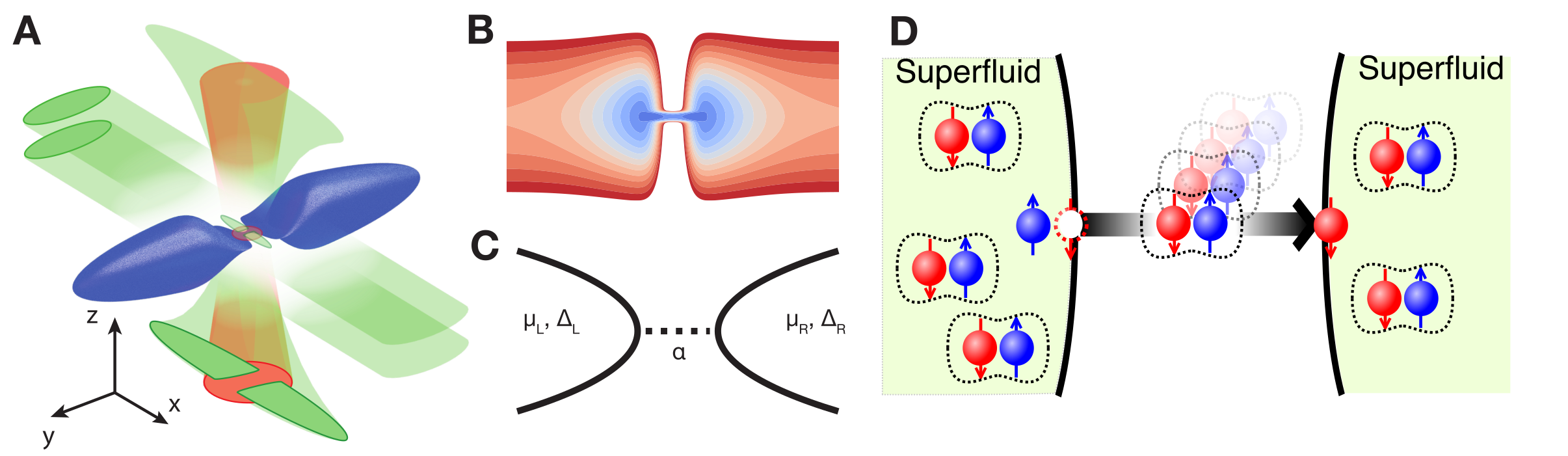}
\caption{{\bf Concept of the experiment}. \textbf{(A)} Schematics of the two 
atom reservoirs (blue) connected by a QPC. Two repulsive beam (green) 
confine the center of the cloud in $x$- and $z$-direction, the attractive 
gate beam (red) tunes the density in the QPC. The dark contours are 
schematics of the beam profiles. \textbf{(B)} Potential landscape in the 
plane $z=0$. Close to the QPC the attractive gate creates areas of high 
density (blue). \textbf{(C)} Theoretical model for the QPC. Both sites of the 
QPC have a defined atom number, imposing a chemical potential $\mu_{\mathrm{L
}}$, $\mu_{\mathrm{R}}$ and a pairing gap $\Delta_{\mathrm{L}}$,$\Delta_{
\mathrm{R}}$. The transparency $\alpha$ is an energy dependent function 
describing the transmission of single particles from one site to the other. 
\textbf{(D)} Transport via multiple Andreev reflections. Coherent tunnelling 
of pairs allows for the creation and tunnelling of a single particle 
excitation (pair breaking) leading to a DC current, even for $\Delta \mu \ll 
\Delta$.}
\label{fig:Fig1}
\end{figure*}

Recent progresses in the manipulation of cold atomic gases have allowed to 
create a mesoscopic device featuring quantised conductance between two 
reservoirs in the non-interacting regime \cite{Krinner:2015aa}. We use this 
technique to create a QPC in a strongly interacting Fermi gas consisting of $1
.7(2)\times 10^5$ $^6$Li atoms in each of the two lowest hyperfine states, in 
a magnetic field of 832 G, where the interaction strength diverges due to a broad Feshbach resonance. The atoms form a strongly correlated 
superfluid, with a pairing gap larger than the chemical potential \cite{
Zwerger:2011aa}. Typical temperatures in the cloud are $T=100(4)$ nK at a 
chemical potential of $\mu=360\ \mathrm{nK}\cdot k_B$. The setup is presented 
in Figure \ref{fig:Fig1}A \cite{materialsandmethods}. The QPC is 
characterised by transverse trapping frequencies of $\nu_x=10.0(4)$ and $\nu_z
=10(3)$ kHz in $x$- and $z$-direction. An optical attractive "gate" 
potential is used to tune the chemical potential and the number of 
channels in the QPC (see Figure~\ref{fig:Fig1}B) \cite{materialsandmethods}. 
We prepare two atomic clouds (reservoirs) with an atom number difference $\Delta N$ while blocking transport through the QPC with a repulsive 
laser beam. This results in a chemical potential bias between the two 
reservoirs $\Delta \mu = f(\Delta N,T/T_{\mathrm{F}})$, with the Fermi temperature $T_F$ and the function $f$ derived from the equation of state (EoS)
of the trapped Fermi gas at unitarity \cite{Ku:2012aa,materialsandmethods}.

\begin{figure*}[htb]
	\centering
		\includegraphics{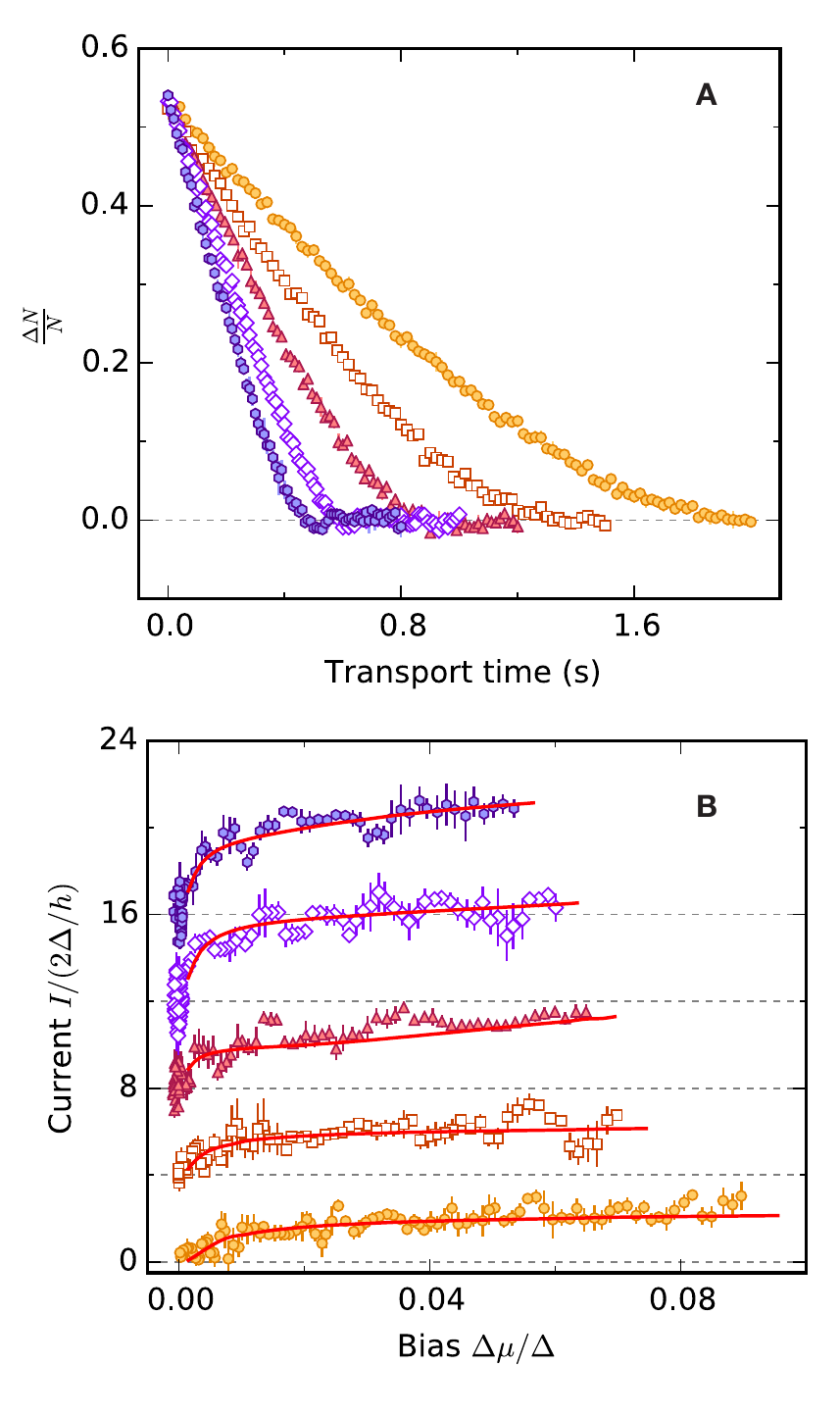}
	\caption{{\bf Non-linear characteristic of the QPC}. \textbf{(A)} Time 
evolution of the particle imbalance $\Delta N
 / N$ for $V_{\mathrm{g}}=$341~nK (filled circle),
 443~nK (open square), 544~nK (filled triangle), 645~nK (open diamond)
 and 747~nK (filled hexagon). \textbf{(B)} Current-bias characteristics
 normalised with respect to $\Delta$. The
 error bars represent the variation of three averaged data
 sets. Negative values of the current are artifacts from the numerical
 derivation process. The red lines show the result of Keldysh
 calculations with the transparency $\alpha$ of the QPC as the only free 
parameter (see \cite{materialsandmethods}). For clarity the curves are 
shifted vertically by 4 units.}
	\label{fig:Fig2}
\end{figure*}

We open the QPC and measure $\Delta N$ as a function of time $t$. Figure~
\ref{fig:Fig2}A presents this evolution for various strengths of the gate 
potential $V_G$. We observe that the $\Delta N$ decays from its initial value 
to zero over a time scale of 0.5 to 1.5s. The shape of the decay curves 
deviates from an exponential and is a direct manifestation of a non-
linear relation between $\Delta N$ and the atom current.

We extract the numerical derivative of these data \cite{materialsandmethods}, 
yielding the instantaneous current at a certain $\Delta \mu$ shown in Figure \ref{fig:Fig2}B. We normalise the 
current and bias by the strength of the pairing gap $\Delta$, using the known relation of $\Delta$ with the chemical potential $\mu$ 
for the low temperature Fermi gas, $\Delta =\eta / \xi \cdot\mu$, with $\eta=0
.44$ \cite{Schirotzek:2008aa} and $\xi=0.37$ \cite{Ku:2012aa,Zurn:2013aa}. 


The current-bias characteristics in Figure \ref{fig:Fig2}B are strongly non-
linear for all the choices of $V_G$, featuring a very strong 
response at low bias.
The high-bias regime approaches a linear dependance with a nonzero intercept on the current axis, marking an 
excess current that depends on the strength of the $V_G$. The current is much 
larger than what is observed for non-interacting atoms \cite{Krinner:2015aa}, 
as observed in earlier measurements on strongly interacting atoms in 
multimode channels \cite{Stadler:2012kx}. 

We model the 
experimental system as two superfluid reservoirs connected by a single 
particle hopping mechanism solely characterised by the transparency $\alpha$ of the
QPC (see Figure~\ref{fig:Fig2}C) \cite{Blonder:1982aa,Averin:1995aa,Cuevas:1996aa,Bolech:2004aa,PhysRevB.71.024517}. 
This model excludes any finite size and geometry-dependent effects,
which we think of as absorbed in $\alpha$. It is motivated by the large
proximity effect in superfluids separated by a ballistic normal barrier
\cite{Zagoskin:1998aa}. Indeed, we expect a coherence length of
$\frac{\hbar v_F}{k_B T}\sim 3\,\mu$m, where $v_F$ is the Fermi
velocity, $k_B$ is Boltzmann's constant and $T$ is the temperature of
the gas. This is comparable to half the length of the channel (5.6 $\mu$m), 
thus we approximate the channel with a point-like connection. 
Using a non-equilibrium Keldysh Green function technique \cite{Kamenev:2011aa}
with mean-field approximation \cite{materialsandmethods} we compare 
theoretical current-bias curves with our data.

Since the pairing gap and the EoS of the unitary gas are
known a priori, the only free parameter in our model is the transparency
$\alpha_n$ for each transverse mode $n$ in the QPC \cite{materialsandmethods}. The solid lines in
Figure~\ref{fig:Fig2}B show the results with the best fits of $\alpha$.
 For the two lowest gate potentials we obtain good agreement
with a single channel model, whereas for higher gate potentials three
channels are required, in agreement with our reference measurement with a 
weakly interacting Fermi gas \cite{Krinner:2015aa}.

The agreement between theory and experiment clarifies the microscopic origin 
of the current. Reflecting the strongly interacting nature of the system, we 
have $\Delta \mu \ll \Delta$. In this regime, a current flow is allowed via 
multiple coherent reflections of quasiparticles between superconducting 
reservoirs, i.e., multiple Andreev reflections, illustrated in Figure~\ref{fig:Fig1}D \cite{Averin:1995aa}. The gap for a single particle transfer can 
be bridged by the simultaneous, coherent transfer of $n$ pairs if $2n \Delta 
\mu >\Delta$, with a probability of order $\alpha^{2n}$. As is seen in 
Figure~\ref{fig:Fig2}B, the drop of current observed at low bias corresponds 
to $\Delta \mu \sim \Delta(1-\alpha)$ \cite{Averin:1995aa}, where the finite 
transparency $\alpha$ suppresses the corresponding Andreev processes. In the very low bias regime, not resolved 
in the experiment, the DC current is exponentially suppressed and an 
oscillating current caused by the energy mismatch between the two reservoirs 
occurs. This averages to zero in the DC limit, and represents an AC Josephson 
current adding to the DC response \cite{materialsandmethods,PhysRevLett.60.416,RevModPhys.74.741,Albiez:2005aa,Ramanathan:2011aa,LeBlanc:2011aa,Jendrzejewski:2014aa}.

\begin{figure*}[htb]
	\centering
		\includegraphics{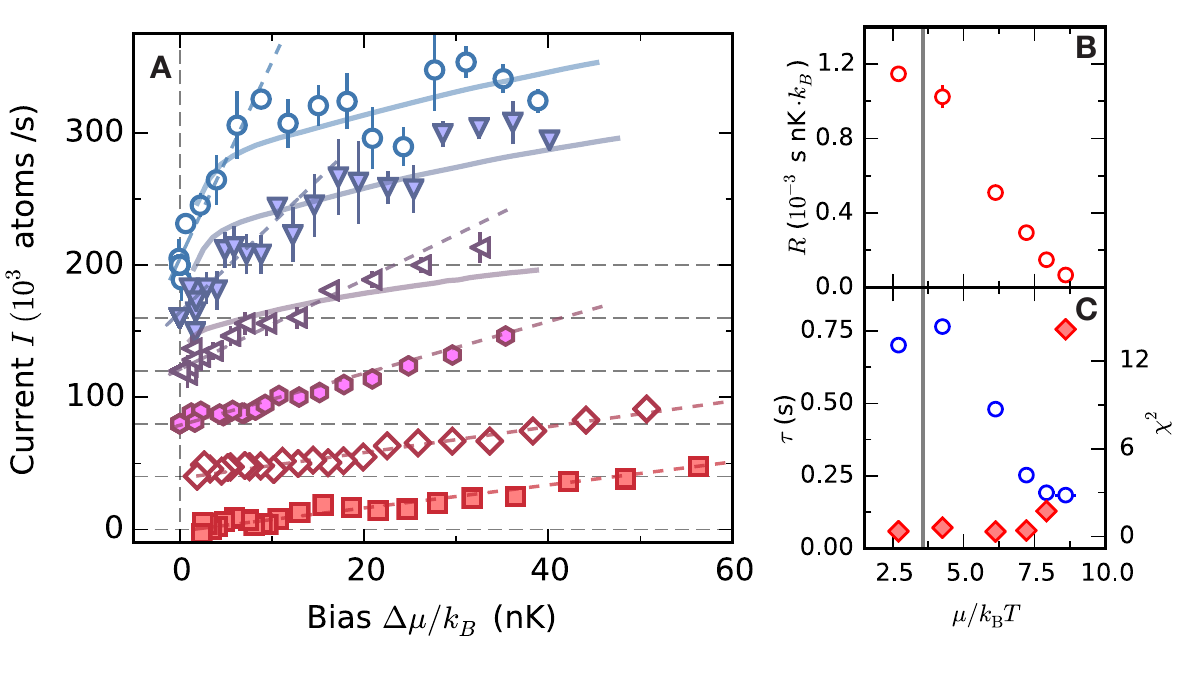}
	\caption{{\bf Finite-temperature transport properties}. \textbf{(A)} Current
-bias curves for temperatures 124 nK (open circle), 135 nK (filled triangle), 
145 nK (open triangle), 170 nK (filled hexagon), 223 nK (empty diamond) and 
290 nK (filled square). For better visibility the curves have been shifted 
vertically by 40 units. Solid lines show best fits of the Keldysh 
calculations. Dashed lines are linear fits to the low bias regime, which 
yields the differential resistance $R=\Delta\mu /I$ shown in \textbf{(B)} as 
a function of decreasing temperature $\frac{\mu}{k_B T}$. \textbf{(C)} Decay 
time $\tau$ of the particle imbalance relaxation (open circle) obtained by 
fitting decay curves with an exponential; $\chi^2$ of the corresponding 
exponential fits (filled diamond) as a function of decreasing temperature $
\frac{\mu}{k_B T}$. The gray solid line marks the superfluid transition at $
\frac{\mu}{k_B T_c}=3.56$}
	\label{fig:Fig3}
\end{figure*}

We now investigate the current-bias relation as a function of temperature, 
for a fixed gate potential $V_G=674$~nK. To this end, we introduce a 
controlled heating of both reservoirs before the transport is started, using 
variable amplitude parametric heating. With this method we explore a 
temperature range of 124-290\,nK, from a deeply superfluid regime up to the 
superfluid-to-normal transition point. We measure the decay of particle 
imbalance with increasing temperature and observe a crossover towards 
exponential decay when temperature is above 145\,nK. We extract the current-
bias characteristic (Figure~\ref{fig:Fig3}A) using the known finite-temperature EoS of 
the unitary Fermi gas \cite{Ku:2012aa,materialsandmethods}. With increasing temperature the non-linearity disappears and the current globally decreases. 
We interpret this as the disappearance of the superfluid contribution to transport as temperature is raised. 


From these data, the differential resistance $R$ at low bias is estimated by 
fitting a line to the low bias region of the curve. The result is presented 
in Figure~\ref{fig:Fig3}B, where the decrease in $R$ is 
clearly visible as temperature is decreased. We compare this to a model 
independent measure of the non-linearity of the characteristic provided by 
the $\chi^2$ parameter of an exponential fit to the entire decay curves \cite{materialsandmethods}. This parameter, as well as the fitted timescale $\tau$ of the 
exponential decay tracks the measured differential resistance (Figure~\ref{fig:Fig3}C). 

While in the low temperature regime, non-linearity is captured by our mean-
field model, the model fails to reproduce the resistance at high 
temperature, indicating a breakdown of the mean-field description.
In particular, it predicts a resistance in the linear regime 
given by the non-interacting Landauer formula, while the resistance that we 
measure is one order of magnitude smaller. This constitutes an indirect 
evidence that the high temperature state of the gas is not a Fermi liquid, 
since Fermi liquid theory leads to the Landauer formula, setting the upper limit for 
the current carried by an ideal contact. 
One possible explanation is the presence of superfluid fluctuations, which 
are expected to be large close to $T_{\mathrm{C}}$ \cite{Liu:2014aa}. Indeed, 
it is known that a one-dimensional Fermi gas held between attractively 
interacting leads, a prototype of non-Fermi liquid system \cite{giamarchi:2003}, shows an enhanced 
conductance \cite{Maslov:1995aa,Safi:1995aa,PhysRevB.52.R8666}. 

\begin{figure*}[htb]
	\centering
		\includegraphics{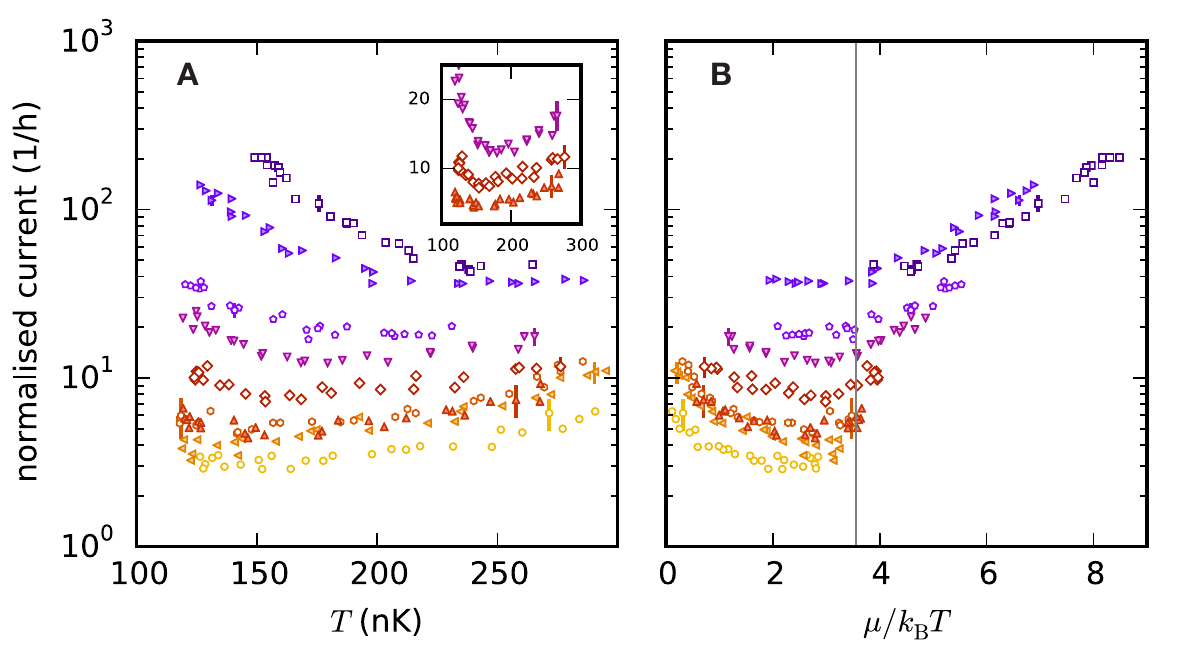}
	\caption{{\bf Competition between superfluidity and thermally activated 
transport}. \textbf{(A)} Normalised current as a function of temperature for 
various $V_G$: 0 nK (open circle), 12 nK (filled triangle left), 54 nK (open 
hexagon), 97 nK (filled triangle up), 139 nK (open diamond), 223 nK (filled 
triangle down), 308 nK (open pentagon), 519 nK (filled triangle right), 942 
nK (open square). Inset: Zoom into the transient region for three selected 
values of $V_G$. \textbf{(B)
} The same data as a function of $\frac{\mu}{k_B T}$. The gray solid line 
marks the superfluid transition at $\frac{\mu}{k_B T_c}=3.56$. Few 
representative error bars are shown.}
	\label{fig:Fig4}
\end{figure*}

The physical picture emerging from these measurements relies on the finite-
temperature properties of the reservoirs. Some of the intrinsic physics of 
the channel appears when the mode spacing in the QPC is comparable to the 
temperature range explored. In this situation, several modes are thermally populated \cite{Brantut:2013aa}, 
enhancing transport for increasing temperature and competing with the 
subsequent reduction of the superfluid current. To show this, we decrease the 
confinement in the QPC along the $x-$direction to $5\,$kHz and systematically 
measure the current at the largest bias as a function of temperature for 
various gate potential. The results are shown in Figure~\ref{fig:Fig4}A 
where the current at high bias normalised to the bias is shown as a function 
of temperature for various gate potentials. In the linear response regime, 
this current-bias-ratio reduces to the differential conductance.

For large $V_G$, we observe a decrease of the current with temperature over 
one order of magnitude. In this superfluid-dominated regime, the pairing gap 
is large compared to both temperature and level spacing in the QPC, analogous to the previous measurements. At low $V_G$, the pairing gap is 
small and vanishes upon heating. In this channel-dominated regime, we observe 
an increase of the normalised current with temperature, which we attribute to 
thermal activation of transport channels in the QPC. There again, the current 
in the high temperature regime is much higher than predicted by a Fermi-
liquid based Landauer formula.

At equilibrium in the reservoirs, superfluidity is universally related to the 
local fugacity \cite{Ho:2004aa} due to the scale invariance of the unitary 
Fermi gas. This suggests the parameter $\mu/k_B T$ as a common 
dimensionless scale for comparing conductances at various temperatures and $V_
G$. Figure \ref{fig:Fig4}B presents the normalised current as a function of $\mu/k_B T$. The data sets showing decreasing current with increasing 
temperature are all grouped in the high fugacity regime, below the expected 
superfluid transition point, confirming that this regime is dominated by 
superfluidity. Conversely, in the low fugacity regime the current increases 
with temperature, corresponding to the channel dominated regime. The 
crossover takes place close to the same fugacity for all the gate potentials, 
and is close to the universal transition point for the unitary Fermi gas at 
the center of the cloud. We expect that the exact location of the crossover, 
as well as the conductance at the minimum depend on the details of the 
channel geometry such as its energy dependent mode spacing. In addition, 
proximity effects should be reduced at high temperature and one dimensional 
physics could emerge in the QPC making the results dependent on its length 
\cite{Buchler:2004aa}. Our setup, allowing for a direct and independent 
control of the geometry, could be used to investigate such effects in future 
experiments.

We acknowledge discussions with Christophe Berthod, Jan von Delft, 
Eugene Demler, Charles Grenier, P\"aivi T\"orm\"a and Johann Blatter. We 
acknowledge financing from NCCR QSIT, the ERC project SQMS, the FP7 project 
SIQS, Swiss NSF under division II and the Ambizione program, the ARO-MURI Non-
equilibrium Many-body Dynamics grant (W911NF-14-1-0003).


\bibliography{paper}
\bibliographystyle{science}

\renewcommand*{\citenumfont}[1]{S#1}
\renewcommand*{\bibnumfmt}[1]{[S#1]}

\newcommand \beq{\begin{eqnarray}}
\newcommand \eeq{\end{eqnarray}}

\onecolumngrid
\pagebreak
\appendix

\section*{Materials and Methods}

\subsection*{Theoretical model}

In general, transport properties of fermionic superfluid
junctions with a few conduction channels can be dealt with in
two different approaches:
scattering formalism based on Bogoliubov-de Gennes equations
\cite{PhysRevB.25.4515,PhysRevLett.75.1831}
and Hamiltonian formalism with nonequilibrium Keldysh Green
function techniques
\cite{PhysRevB.54.7366,PhysRevLett.92.127001,PhysRevB.71.024517}.
We adopt the latter approach, which
allows us to compute currents with high accuracy and potentially to treat
more general systems with interaction and disorder.

As discussed in the main text, since the junction is short,
we approximate the 1D channels connecting two reservoirs by
a quantum point contact. To be specific, we consider the following Hamiltonian:
\beq
&&H=H_L+H_R+H_{t},\\
&&H_{t}=-\sum_{n}\sum_{\sigma=\uparrow,\downarrow}
t_n\psi^{\dagger}_{L,\sigma}(0)\psi_{R,
\sigma}(0)+h.c.
\label{eq:hamiltonian}
\eeq
Here, $H_j$ $(j=L \ \text{or}\ R)$
is the Hamiltonian of the reservoir
$j$ and is idential to the so-called single channel model
\cite{RevModPhys.80.885},
which consists of the quadratic part in the fermionic field $\psi$ including
the kinetic energy, trapping potential, and chemical potential
terms and the quartic part in $\psi$
describing the local pair interaction.
On the other hand, $H_{t}$ describes tunneling processes between
the reservoirs occurring at a single point denoted $x=0$.
The physical properties of the contact are assumed to
be hidden in the tunneling amplitudes $t_n$ of each transverse mode
of the 1D contact $n$ participating the tunneling.
By assuming the above Hamiltonian and a steady state solution, one can
obtain the current as follows:
\beq
I=\frac{1}{2}\langle\frac{\partial(N_R-N_L)}{\partial t}
\rangle=\sum_{n\sigma}\frac{it_n}{\hbar}\langle
\psi^{\dagger}_{L,\sigma}(0)\psi_{R,\sigma}(0)\rangle+h.c.,
\eeq
where $N_R$ and $N_L$ are the total number of particles in the reservoirs and 
the average $\langle \cdots\rangle$ is expressed in terms of
the Keldysh Green functions as discussed below.
An important observation is the fact that the current depends only
on the tunneling terms at position $x=0$ and therefore one can
obtain a local action by
integrating out the position dependence of the reservoir Hamiltonians.

While the above tunneling Hamiltonian has a simple structure,
it is nevertheless difficult to obtain the current
based on it since the Hamiltonian in each reservoir is quartic
in $\psi$.
To go further, we adopt a quadratic (mean-field)
approximation,
 which renders the local Hamiltonian quadratic in $\psi$ as
\beq
{\cal H}_j=\sum_{\sigma}\psi^{\dagger}_{j,\sigma}\left(-\frac{\hbar^2\nabla^2}{2m}
-\mu_j\right)\psi_{j,\sigma}
-\Delta_j\psi^{\dagger}_{j,\uparrow}\psi^{\dagger}_{j,\downarrow}
-\Delta_j\psi_{j,\downarrow}
\psi_{j,\uparrow}.
\eeq
Here, $\mu_j$ and $\Delta_j$ are the chemical potential
and fermionic superfluid gap of the reservoir $j$
defined locally.
This approximation allows one to obtain
 the local retarded and
advanced Green functions in each reservoir
as follows
\cite{PhysRevB.54.7366,PhysRevLett.92.127001,PhysRevB.71.024517}:
\beq
g^{r,a}_j(\omega)=\frac{1}{\sqrt{\Delta_j^2-(\omega-\mu_i\pm i\eta)^2}}
\begin{pmatrix}
-(\omega-\mu_j\pm i\eta) & \Delta_j\\
\Delta_j & -(\omega-\mu_j\pm i\eta)
\end{pmatrix},
\eeq
where we introduce the Nambu representation
for the fermions in each reservoir,
$\eta=0^+$ is an infinitesimal positive parameter that regularizes
the Green functions, and the upper (lower) sign corresponds to
the retarded (advanced) Green function, $g^r$ $(g^a)$.
The Green functions are measured
in units of an energy scale associated with the normal density of states at
the Fermi level
\cite{PhysRevB.54.7366,PhysRevLett.92.127001,PhysRevB.71.024517}.
One  obtains the Keldysh component of the Green functions
by means of the fluctuation-dissipation theorem \cite{kamenev2011field}:
$g^k_j=(g^r_j-g^a_j)\tanh((\omega-\mu_j)/2T)$ with the temperature $T$.
By using the Green functions obtained above,
the corresponding local action can be expressed as
\beq
&& S=\int \frac{d\omega}{2\pi}\sum_{j=L, R}
\Psi^{\dagger}_jg^{-1}_j\Psi_j
+S_{t},
\label{eq:fullaction}\\
&& S_{t}=-\int \frac{d\omega}{2\pi}
\sum_{n,\sigma}t_n\psi^{\dagger}_{L,\sigma}(\omega)\psi_{R,\sigma}
(\omega)+h.c.
\eeq
Note that $\Psi$ is a four components spinor consisting of the Keldysh and Nambu
components, and the matrix $g^{-1}_j$ is given by
\beq
g^{-1}_j=
\begin{pmatrix}
(g^{r}_j)^{-1} & (g^k_j)^{-1}\\
0 & (g^{a}_j)^{-1}
\end{pmatrix},
\eeq
where
$(g^k_j)^{-1}=-(g^{r}_j)^{-1}g^k_j(g^{a}_j)^{-1}$
\cite{kamenev2011field}.

The action (\ref{eq:fullaction}) and the two reservoirs and the tunnelling term could be regrouped in an 
eight by eight matrix. It is in principle possible to invert it and thus to obtain the full solution of this problem.
Note however that frequencies with equal positive and negative shifts from
the Fermi level ($\pm (\omega-\mu_j)$) are used in the Nambu representation
while absolute frequencies $\omega$ are conserved by
the tunneling term ($S_t$).
This implies that, even if it is quadratic, in the presence of a finite bias
the action is no longer diagonal in frequencies and
one must consider an infinite dimensional matrix to be inverted.
However, only a discrete set of frequencies are coupled by the tunnelling term. 
Physically, such an infinite discrete set of frequencies represents
multiple Andreev reflections (See Figure 1 D in the main text and Refs.~\cite{PhysRevB.25.4515,PhysRevLett.75.1831,PhysRevB.54.7366,PhysRevLett.92.127001,PhysRevB.71.024517}), which are responsible for the 
non-linear $I-V$ behavior in fermionic superfluids.
Since the weight of the terms with frequencies far away from the Fermi levels
decreases, it is possible to truncate the infinite series
of the multiple Andreev
reflections \cite{PhysRevLett.92.127001,PhysRevB.71.024517}
and thus to invert numerically the matrix with the desired accuracy.

Inverting the matrix allows to compute the current. Indeed the current can be expressed as 
\cite{PhysRevB.54.7366,PhysRevLett.92.127001,PhysRevB.71.024517}
\beq
I=\sum_{n}\frac{2t_n}{h}\int d\omega \text{Re}G^{K},
\eeq
where $G^K$ is the Keldysh component appearing in the inverse of the full action 
\cite{PhysRevB.54.7366,PhysRevLett.92.127001,PhysRevB.71.024517}.

If both reservoirs are normal ($\Delta_j=0$),
the offdiagonal elements in the Nambu representation disappear
and the local action is diagonal in frequencies.
One can then compute analytically the current and obtains
the Landauer-B\"uttiker formula \cite{landauer,PhysRevLett.57.1761}:
\beq
I=\sum_n\frac{2\alpha_n}{h} \int d\omega [f_L(\omega)-f_R(\omega)],
\label{eq:landauer}
\eeq
where $f_j$ is the Fermi distribution function in the reservoir $j$,
and $\alpha_n$ is the so-called transparency associated
with the hopping amplitude $t_n$ as
\beq
\alpha_n=\frac{4t_n^2}{(1+t_n^2)^2}.
\eeq
The transparency takes values in the interval $[0,1]$,
and perfectly transparent junctions corresponds to $\alpha=1$.

Using the above formalism, one can thus compute the current for the nonzero
$\Delta$  case as well by fixing the hopping amplitudes (see Figure~\ref{fig:hopings}). Results are given in Figure 2 B of the main text.
In the case of the experimental system considered in the present paper,
the junction is near perfect transparency $\alpha\approx1$.

\subsection*{Data fitting}

\begin{figure}
\def \thefigure{S1}
	\centering
		\includegraphics{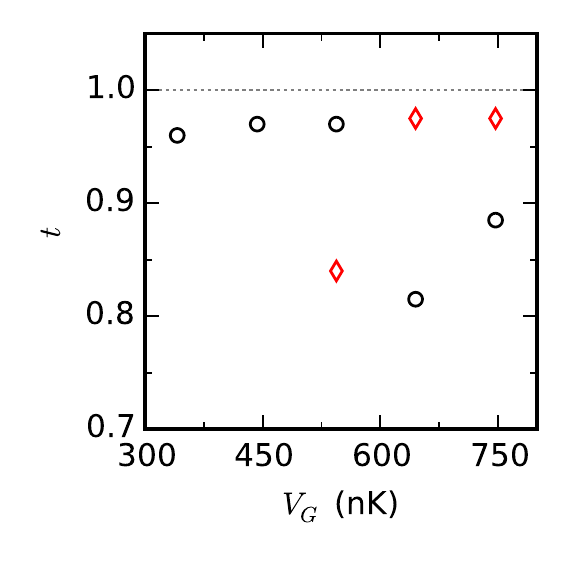}
		\caption{Tunnelling amplitudes fitted at low temperature as a function of gate potential. For the lowest gate potentials, a single tunnelling parameter is fitted for the lowest mode (black rings). For higher gate potentials, three amplitudes are fitted, one for the lowest mode and two identical for the next two quasi-degenerate modes (red diamonds).}
	\label{fig:hopings}
\end{figure}

The data presented in Figure 2B are fitted using the results of the numerical calculation of the current-bias characteristic, leaving only $t_n$, the tunnelling amplitudes, as fit parameters. For each choice of gate potential the parameters are allowed to vary, reflecting the fact that the transparency is affected by the adiabaticity of the coupling and by the change of geometry induced by the gate potential. Figure \ref{fig:hopings} presents the fitted tunnelling amplitudes as a function of the gate potential. The lowest fitted value is $\sim 0.8$, leading to a transparency for the corresponding mode larger than $0.95$, compatible with the observation of quantized conductance in the non-interacting regime. Most of the fitted values are $>0.95$, corresponding to a transparency $\alpha>0.99$.

\subsection*{Thermodynamics}

We make use of the known thermodynamic relations for harmonically trapped and homogeneous unitary Fermi gases to extract the chemical potential bias and temperature of the gases. In this section we describe the employed procedure. 

\subsubsection*{Equation of states}
The density equation of state (EoS) of the homogeneous unitary Fermi gas can be expressed as a universal function of $q=\mu / k_B T$ \cite{ku2012revealing}
\begin{equation}
	n\lambda^3 = f_n(q) = -\mathrm{Li}_{3/2}\left(-e^{q}\right)F(q)
\label{eq:fn}
\end{equation}
with the de Broglie wavelength $\lambda=\sqrt{\frac{2\pi\hbar^2}{mk_b T}}$. Here the polylogarithm function represents the noninteracting part, while $F(q)$ is the ratio $n(q)/n_0(q)$ between the unitary and noninteracting density as measured in \cite{ku2012revealing} for $-0.9<q<3.5$. Using a fourth order virial expansion we extend the EoS to $q<-0.9$ \cite{ku2012revealing,PhysRevLett.102.160401}. Further, we use the phonon model as described in \cite{taylor_first_2009,PhysRevA.88.043630} to expand the EoS to $q>3.5$. The full EoS is then given by:
\begin{equation}
	f_n(q)=
	\begin{cases}
		\sum_{i}^{j}b_j j e^{iq},\ q<-0.9\\
		F_n(q)\left(-\mathrm{Li}_{3/2}(-e^{q})\right),\ -0.9<q<3.5\\
		\frac{(4\pi)^{3/2}}{6\pi^2}\left[\left(\frac{q}{\xi}\right)^{3/2} -\frac{\pi^4}{480}\cdot\left(\frac{3}{q}\right)^{5/2} \right],\ 3.5<q
	\end{cases}
	\label{eq:Eos}
\end{equation}

In the local density approximation (LDA), each point in the cloud has a local chemical potential given by
\begin{equation}
	q(x,y,z)=\frac{\mu(x,y,z)}{k_B T} = \frac{1}{k_B T}\left(\mu_0 - \frac{1}{2}m\left(\omega_x^2 x^2 + \omega_y^2 y^2 + \omega_z^2 z^2\right)\right)
\label{eq:Vxyz}
\end{equation}
with the chemical potential $\mu_0$ in the bottom of the trap.
The total atom number $N$ in LDA is then given by the density integral over the full cloud
\begin{equation}
	N=\frac{1}{\lambda^3}\int dxdydz\ f_n \left(q(x,y,z)\right)= \frac{2}{\sqrt{\pi}}\left(\frac{k_B T}{\hbar\bar{\omega}}\right)^3 M_0(q_0)
\label{eq:Ntot}
\end{equation}
with the average trapping frequency $\bar{\omega}$ and the dimensionless moments \cite{PhysRevA.87.063601}
\begin{equation}
	M_l = \int_{-\infty}^{q_0}dq \left(q_0 - q \right)^{(l+1)/2} f_n(q)
\label{eq:Ml}
\end{equation}

Using the known expression for the Fermi temperature $k_B T_F=\hbar\bar{\omega} (6N)^{1/3}$ we obtain the degeneracy factor
\begin{equation}
	\frac{T}{T_{\mathrm{F}}} = \left( \frac{12}{\sqrt{\pi}}M_0 (q_0) \right)^{-\frac{1}{3}} =: h(q_0)
\label{eq:TTf_vs_mu}
\end{equation}
where the function $h(q_0)$ formally represents the dependance of $T/T_F$ on $q_0$. We now invert $h(q_0)$ and obtain the universal function $g$:
\begin{equation}
	q_0=\frac{\mu_0}{k_B T}=g\left(\frac{T}{T_F}\right) = h^{-1}\left(\frac{T}{T_F}\right)
\label{eq:betamuVsT}
\end{equation}
Since we have determined $T$ independantly, the function $g$ directly yields the global chemical potential in the harmonic trap $\mu_0=g(T/T_F)\cdot k_B T$ in the reservoirs as a function of temperature.

In a similar way we use the EoS to convert an atom number imbalance $\Delta N$ into a chemical potential bias $\Delta\mu$ accounting for finite temperature. To this end we take the logarithm of Eq.~\eqref{eq:betamuVsT} and derive it with respect to the atom number $N$
\beq
	\frac{d}{dN}\log(\mu) = \frac{d}{dN}\log\left(k_B T g\left(\frac{T}{T_F}\right)\right)\\
	\frac{1}{\mu}\frac{d\mu}{dN} = -\frac{1}{3}\frac{g'\left(\frac{T}{T_F}\right)}{g\left(\frac{T}{T_F}\right)}\frac{T}{T_F} \frac{1}{N}
\label{eq:deriveMu}
\eeq
where we have denoted the derivative of $g$ as $g'$. Using Eq.~\eqref{eq:betamuVsT} to eliminate $\mu$ and making a first order approximation in $\Delta N / N$ in Eq.~\eqref{eq:deriveMu}, we obtain an expression for the chemical potential bias $\Delta\mu$ as a function $\Delta N$, $N$ and $T/T_F$

\begin{equation}
	\frac{\Delta\mu}{E_F} = -\frac{1}{3} g' \left(\frac{T}{T_F}\right)^2 \frac{\Delta N}{N}
\label{eq:deltaMu}
\end{equation}

\subsubsection*{Finite temperature compressibility}
\begin{figure}
\def \thefigure{S2}
	\centering
		\includegraphics{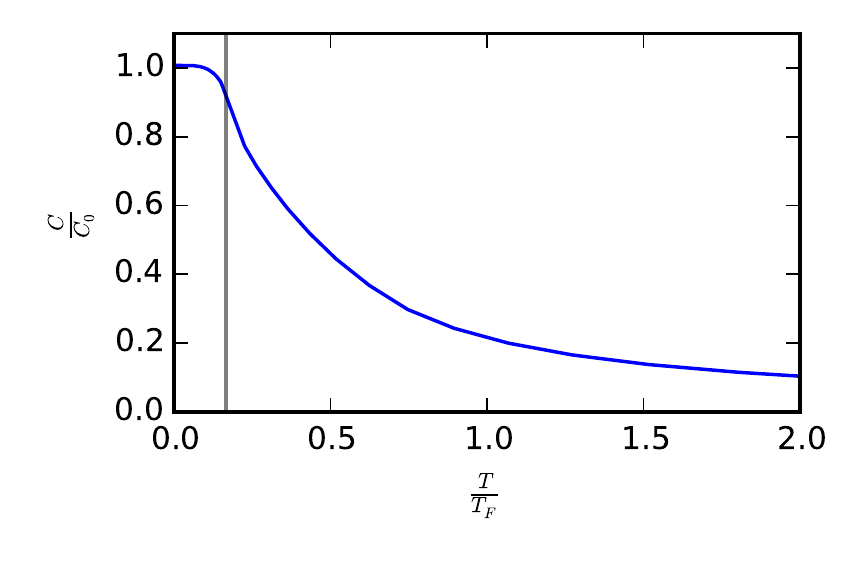}
		\caption{Ratio $C/C_0$ as a function of $T/T_F$. The veritcal line indicates the critical temperature at $\mu / k_B T=2.5$}
	\label{fig:Compressibility}
\end{figure}

The finite temperature compressibility at unitarity is given by the derivative of Eq.~\eqref{eq:Ntot}: 
\begin{equation}
	C = \frac{\partial N}{\partial\mu} = \frac{1}{2\sqrt{\pi}}\frac{(k_B T)^2}{(\hbar\bar{\omega})^3}M_{-2}(q_0)
\label{eq:C}
\end{equation}
Comparing this to the zero-temperature limit compressibility
\begin{equation}
	C_0 = \frac{\partial N}{\partial\mu}^0 = \frac{1}{2}\frac{(6N)^{3/2}}{\sqrt{\xi}\hbar\bar{\omega}}
\label{eq:C0}
\end{equation}
we find a ratio between $C$ and $C_0$ of
\begin{equation}
	\frac{C}{C_0} = \frac{\sqrt{\xi}}{12^{2/3}\pi^{1/6}}\frac{M_{-2}\left(g(T/T_F)\right)}{M_1 \left(g(T/T_F)\right)^{2/3}}
\label{eq:CratioC0}
\end{equation}
which is plotted in Fig.~\ref{fig:Compressibility}.

\subsubsection*{Temperature calibration}
The temperature of the unitary Fermi gas can be obtained using the relation between the second moment of the density $\left<y^2\right>$ and the total energy per particle $E/N$ \cite{PhysRevLett.95.120402}
\begin{equation}
	E=3m\omega_y\left<y^2\right>
\label{eq:virial}
\end{equation}
with the second moment
\begin{equation}
	\left<y^2\right>=\int_{-\infty}^{\infty}dy n_{1D}(y)y^2
\label{eq:2ndMoment}
\end{equation}
We select the direction $y$, because the cloud expands very slowly in $y$ during the time of flight of 0.8 s, due to the magnetic confinement in $y$. The density profile $n_{1D}(y)$ after time of flight is then indistinguishable from in situ pictures. Further, we use the EoS (Eq.~\eqref{eq:fn}) to express the total energy per particle with reference to the Fermi level $E/NE_F$. To this end we integrate the energy density $e=n\cdot(q_0-q)$ over the full cloud and obtain
\begin{equation}
	\frac{E}{NE_F}	= \left(\frac{2\pi}{9}\right)^{\frac{1}{6}}\frac{M_2(q_0)}{M_0^{4/3}(q_0)}
\label{eq:energyDensity}
\end{equation}
We can now combine Eq.~\eqref{eq:energyDensity} and Eq.~\eqref{eq:virial} to obtain $q_0$. Plugging this $q_0$ into Eq.~\eqref{eq:TTf_vs_mu} yields the temperature $T$.

Between the end of the transport process and the moment the cloud is imaged in the experiment cycle, the cloud is heated due to spontaneous emission and parasitic heating, with a heating rate $R$ dependant on the power of the dipole trap laser. To obtain the temperature during transport $T$, we thus subtract the total heating in the dipole trap from the temperature at imaging $T_{\mathrm{im}}$.
\begin{equation}
	T = T_{\mathrm{im}} - R t                                                                              
\label{eq:dipoleHeating}
\end{equation}
where $t$ is the time span between the mean transport time and the imaging. In the data of Figure~2 the dipole trap was decompressed during transport and recompressed for imaging, which leads to a decreased heating rate. Furthermore the decompressed cloud has a $E_F$ that differs from the one during imaging due to the dependancy on $\bar{\omega}$. For an adiabatic recompression $T/T_F$ stays constant, so the temperature   scales with the ratio between the trap frequency of the compressed ($\bar{\omega_c}$) and decompressed ($\bar{\omega_d}$) cloud. Summarising all these corrections then yields the temperature during transport
\begin{equation}
	T = \frac{\bar{\omega}_d}{\bar{\omega}_c}\cdot \left(T_{\mathrm{im}} - R_c t_c \right) -R_d t_d
\label{eq:temperatureFinal}
\end{equation}
with the waiting time $t_c$ ($t_d$) and heating rate $R_c$ ($R_d$) in the compressed (decompressed) trap.

\subsubsection*{Chemical potential in the point contact}

Determining the chemical potential using the equation of state implies the assumption of and ideal harmonic trap. However the more complex gemoetry at the center of the cloud in the underlying potential of our system due to the gate potential and the confinement beams in $x-$ and $z-$direction must be taken into account. We corrected for two effects in estimating the chemical potential in the point contact:
\begin{enumerate}
\item The first correction originates from the fact that the QPC confinement in $z$ expels atoms from the center of the cloud, leading to an increased global chemical potential. This confinement is absent during imaging, thus it must be corrected for explicitly. The confinement pushes atoms into the reservoirs, increasing the chemical potential $\mu_{conf}$ with respect to the harmonic trap case $\mu$. The correction factor $b = \mu_{conf}/\mu$ is dependant on the trapping frequency of the harmonic trap $\bar{\omega}$ and the total atom number $N$. We estimate $b$ by computing the density distribution using the Thomas-Fermi approximation in the full trap based on the homogeneous equation of state. For the measurements presented here we find correction factors $b$ between 1.21 and 1.27. 
\item A second correction to the chemical potential arises from finite darkness ($ \sim 99.8\%$) of the same repulsive beam in the QPC, leading to a n additional repulsive optical potential in the nodal line where atoms reside, with an amplitude $V_{fd}=140\ \mathrm{nK}$ and a gaussian waist of $w_{fd}=30\ \mu$m. This is confirmed by a measurement of the conductance of non interacting Fermions in the point contact, compared with an {\it ab-initio} calculation with the Landauer formula.
\end{enumerate}
The chemical potential and the gap are evaluated in the trap center under disregard of confinement induced effects. We are presentaly not aware of calculations of the gap and chemical potential in the presence of 3D scattering and weak confinement.

\subsection*{Numerical derivative of decay curves}
To obtain the current-bias curves shown in the main text, we calculate the numerical derivative of the measured $\Delta N$ with respect to time. To this end we extract the local slope using a linear fit over a window of six consecutive points. We then shift the window by one point and repeat. Note that the overlap of windows creates correlations between adjacent point in the derivative, which leads to oscillating artefacts in the current bias curves.

From the local slope $d\Delta N/d t$ one obtains the current $I$:
\begin{equation}
	I= \frac{d N_L}{dt} = -\frac{d N_R}{dt} = \frac{1}{2} \frac{d (N_L - N_R)}{dt} =\frac{1}{2}\frac{d\Delta N}{d t}
\label{eq:current}
\end{equation}

\subsection*{Exponential fits}
\begin{figure}
\def \thefigure{S3}
	\centering
		\includegraphics{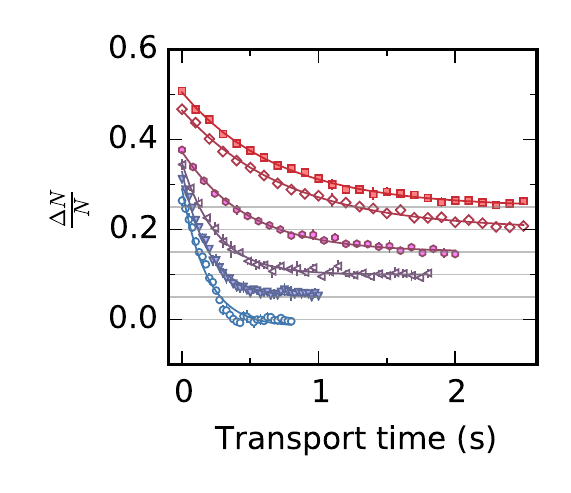}
		\caption{Decay curves for various temperatures: $T=124$ nK (open circle), 135 nK (filled triangle), 145 nK (open triangle), 170 nK (filled hexagon), 223 nK (empty diamond) and 290 nK (filled square). The solid lines are exponential decay fits (see Eq.~\eqref{eq:expfit}). The data are offset by 0.05 units vertically for clarity.}
	\label{fig:tempDecays}
\end{figure}

We fit the decay curves at various temperatures (see Figure~\ref{fig:tempDecays}) with an exponential fit of the from
\begin{equation}
	\frac{\Delta N}{N} (t) = n(t)= n_0 e^{-\frac{t}{\tau}}
\label{eq:expfit}
\end{equation}
with the initial imbalance $n_0$ and the decay time $\tau$ as fit parameters. To test the merit of the exponential fit, we deduce the deviation between the fit and experiemental data. For each decay curve only values for $t<3\cdot \tau$ are taken into account, and the obtained value $\chi^2$ is divided by the number of points $p$:
\begin{equation}
	\chi^2 = \frac{1}{p}\sum_{i,t_i<3\tau}{\frac{\left((\Delta N/N)_i - n(t_i)\right)^2}{\sigma_i^2}}
\label{eq:chi2}
\end{equation}
Here $(\Delta N/N)_i$ is the imbalance form the data at $t_i$, $\sigma_i$ is the standard deviation from averaging in $\Delta N/N_i$ and $n(t_i)$ is the imbalance as obtained from the fit at $t_i$.

\subsection*{Current determination}
To obtain the normalized current shown in Figure 4 we compare the initial particle imbalance $\Delta N_0$ to the imbalance $\Delta N$ after a short time $\Delta t$, determined for each value of the gate potential separately. The mean current $\bar{I}$ in this time window is then given by $\bar{I} = \frac{\Delta N_0-\Delta N}{\Delta t}$. Normalizing this with respect to the chemical potential bias $\Delta \mu$ yields 
\begin{equation}
	\frac{\bar{I}}{\Delta \mu} = \frac{C\bar{I}}{\Delta \bar{N}}
\label{eq:normCurrent}
\end{equation} 
which is plotted in Figure 4. Here we made a linear approximation in the compressibility $C=\Delta \bar{N} / \Delta \mu$, and used the harmonic mean $\Delta \bar{N}$ between $\Delta N$ and $\Delta N_0$ to account for averaging over a finite range of the bias.

\newpage

\end{document}